# The Gravitational Mass at the Superconducting State

Fran De Aquino   physics/0201058

Maranhao State University,
Physics Department,
65058-970 S.Luis/MA, Brazil.
E-mail: deaquino@elo.com.br

## Abstract

It will be shown that the gravitational masses of the electrons of a superconducting material are strongly *negative*. Particularly, for an amount of mercury ($Hg$) at the transition temperature, $T_c = 4.15$ K, the negative gravitational masses of the electrons decrease the *total* gravitational mass of the $Hg$ of approximately 0.1 percent. The weight reduction *increase* when the $Hg$ is spinning inside a magnetic field or when it is placed into a strong oscillating EM field.

## Introduction

We have shown in a previous paper[1] that the *gravitational mass* and the *inertial mass* are correlated by a dimensionless factor, which depends on the incident *radiation* upon the particle. It was shown that only in the absence of electromagnetic radiation this factor becomes equal to 1 and that, in specific electromagnetic conditions, it could be reduced, nullified or made negative.

The general expression of correlation between gravitational mass $m_g$ and inertial mass $m_i$, is given by

$$m_g = m_i - 2\left\{\sqrt{1+\left\{\frac{q}{m_i c}\right\}^2} - 1\right\}m_i \qquad (1)$$

the *momentum* $q$ is given by

$$q = N\hbar k = N\hbar\omega/(\omega/k) = U/(dz/dt) =$$
$$= U/v \qquad (2)$$

where $U$ is the electromagnetic energy absorbed (or emitted) by the particle; $v$ is the velocity of the incident (or emitted) radiation, which is

$$v = \frac{c}{\sqrt{\frac{\varepsilon_r \mu_r}{2}\left(\sqrt{1+(\sigma/\omega\varepsilon)^2}+1\right)}} \qquad (3)$$

where $\omega = 2\pi f$ ; $f$ is the frequency of the radiation: $\varepsilon, \mu$ and $\sigma$, are the electromagnetic characteristics of the outside medium around the particle in which the incident (or emitted) radiation is propagating ($\varepsilon = \varepsilon_r \varepsilon_0$; $\varepsilon_r$ is the *relative electric permittivity* and $\varepsilon_0 = 8.854 \times 10^{-12} F/m$ : $\mu = \mu_r \mu_0$; $\mu_r$ is the *relative magnetic permeability* and $\mu_0 = 4\pi \times 10^{-7} H/m$).

The general expression of correlation between gravitational mass and inertial mass ( Eq.(1) ) was experimentally confirmed by an experiment using Extra-Low Frequency (ELF) radiation on ferromagnetic material. The experimental setup and the obtained results were presented in a previous paper[2]. Recently another experiment[3] using UV light on phosphorescent plastic have also confirmed the Eq.(1).

By the substitution of Eqs.(3) and (2) into Eq.(1), we obtain

$$m_g = m_i - 2\left\{\sqrt{1+\left[\frac{U}{m_i c^2}\sqrt{\frac{\varepsilon_r \mu_r}{2}\left(\sqrt{1+(\sigma/\omega\varepsilon)^2}+1\right)}\right]^2}-1\right\}m_i$$

$$= m_i - 2\left\{\sqrt{1+\left[\frac{U}{m_i c^2}n_r\right]^2}-1\right\}m_i \quad (4)$$

In the equation above, $n_r$ is the *refractive index*, which is given by:

$$n_r = \frac{c}{v} = \sqrt{\frac{\varepsilon_r \mu_r}{2}\left(\sqrt{1+(\sigma/\omega\varepsilon)^2}+1\right)} \quad (5)$$

$c$ is the speed in vacuum and $v$ is the speed in medium.

It is important to note that the electromagnetic characteristics, $\varepsilon$, $\mu$ and $\sigma$, do not refer to the particle, but to the outside medium around the particle in which the incident ( or emitted) radiation is propagating. For an *electron* inside a body, the incident (or emitted) radiation on this electron will be propagating inside the body, and consequently, $\sigma = \sigma_{body}$, $\varepsilon = \varepsilon_{body}$, $\mu = \mu_{body}$. Thus, according to the Eq.(4), the gravitational mass of the electron is given by

$$m_{ge} = m_e - 2\left\{\sqrt{1+\left\{\frac{U}{m_e c^2}n_{r(body)}\right\}^2}-1\right\}m_e \quad (6)$$

where $m_e$ is the *inertial* mass of the electron and $n_{r(body)}$ is the index of refraction of the body.

Based on the equation above, we will be show that the *gravitational masses* of the electrons of a superconducting material are strongly negative. Particularly, for an amount of mercury (*Hg*) at the transition temperature, the negative gravitational masses of the electrons decrease the *total gravitational mass* of the *Hg* of approximately 0.1%.

## 2. Superconductors

Usually, for *superconducting materials*, we have $\sigma \gg \omega\varepsilon$. Thus, in that case, the index of refraction given by the Eq.(5), can be written

$$n_r = \sqrt{\frac{\mu\sigma c^2}{2\omega}} \quad (7)$$

The equation above shows that the *refractive indices* of the superconducting materials are enormous. Consequently, in agreement with Eq.(6), the gravitational masses of their electrons can be significantly reduced even for $U$ relatively small as, for example, in the case of *thermal radiation* emitted from a disk of superconducting material at the transition temperature $T_c$.

In the case of *thermal radiation*, it is common to relate the energy of photons to *temperature*, $T$, through the relation,

$$\langle hf \rangle \approx \kappa T$$

where $\kappa = 1.38 \times 10^{-23}$J/K is the Boltzmann's constant. Thus, we can write

$$U = \eta\langle hf \rangle \approx \eta\kappa T.$$

Where $\eta$ is a particle-dependent absorption ( or emission ) coefficient. Consequently, we can express Eq.(6) as follows:

$$m_{ge} = m_e - 2\left\{\sqrt{1+\left\{\frac{\eta_e}{m_e c}\sqrt{\frac{\mu\sigma\hbar\kappa T}{2}}\right\}^2}-1\right\}m_e \quad (8)$$

For electrons[4] $\eta = \eta_e \cong 0.1$. Thus, for $T = T_c$ (Transition Temperature) the equation above can be written

$$m_{ge} = m_e - 2\left\{\sqrt{1+1.2\times10^{-22}\sigma T_c} -1\right\}m_e \qquad (9)$$

From this equation we can conclude that *the gravitational masses* of the electrons of a superconducting material ($\sigma > 10^{22} S/m$) are strongly *negative*.

Let us now consider the $Hg$ at the *transition temperature*[5], $T_c = 4.15$ K. At this temperature, the electric conductivity of the $Hg$ is $\sigma \cong 10^{23} S/m$. Consequently, from the Eq.(9) we obtain

$$m_{ge} \cong -11.25 m_e$$

Thus the *total* gravitational mass of the $Hg$ is

$$m_{g(Hg)} = \left[\frac{-11.25m_e + m_p + m_n}{m_e + m_p + m_n}\right] m_{i(Hg)} \cong 0.99 m_{i(Hg)}$$

This means that the negative gravitational masses of the electrons decrease the *total gravitational mass* of the $Hg$ of *less than* 1% percent ( approximately 0.1% ).

This prediction can be verified in a very simple way, see Fig. 1. In addition, we suggest to check the weights of samples (of different masses and chemical compositions) hung above the $Hg$. Possibly the percentage of weight decrease will be *the same* of the $Hg$ up to some meters upwards( due to shielding effect produced to the reduction of the gravitational mass of the $Hg$ ).

It is important to note that, if the $Hg$ cylinder *rotates* at a strong magnetic field $B$ perpendicular to the cylinder, the weight reduction must *increase* due to the emission of radiation from the electrons rotating within the magnetic field. At this case, the radiation emitted from each electron has power $P$ which, as we know, is given by[6]

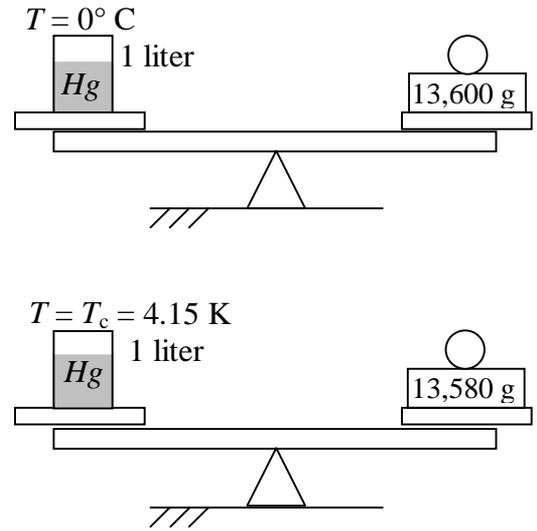

Fig.1 - The Gravitational Mass of the $Hg$ at 0°C and at 4.15K( Transition Temperature)

$$P = \frac{\mu e^4 B^2 V^2}{6\pi m_e^2 c\left(1 - V^2/c^2\right)} \qquad (10)$$

Most of the emitted radiation has frequency $f$

$$f = f_0 \left(1 - \frac{V^2}{c^2}\right)^{-3/2} \qquad (11)$$

where

$$f_0 = \frac{eB}{2\pi m_e}\sqrt{1-V^2/c^2} \qquad (12)$$

is the named *cyclotron frequency*.

To simplify the calculations we can consider only the contribution of the emitted radiation with frequency $f$. Then we can put $U = nhf$ in Eq.(6), where $n$ is the number of emitted

photons from the electron. Thus we can write

$$m_g = m_e - 2\left\{\sqrt{1+\left\{\frac{nhf}{m_e c^2}n_r\right\}^2} - 1\right\}m_e \quad (13)$$

But $P = nhf/\Delta t = nhf^2$, thus we can write $n = P/hf^2$. Substitution of $n$ into Eq.(13) gives

$$m_g = m_e - 2\left\{\sqrt{1+\left\{\frac{P}{m_e cvf}\right\}^2} - 1\right\}m_e \quad (14)$$

For $\sigma >> \omega\varepsilon$ Eq.(3) reduces to

$$v = \sqrt{\frac{4\pi f}{\mu\sigma}} \quad (15)$$

By the substitution of Eqs.(10) and Eq.(15) into Eq.(14) we obtain

$$m_g = m_e - 2\left\{\sqrt{1+\left\{\frac{e^4 B^2 V^2}{6\pi n_e^3 c^2 (1-V^2/c^2)}\sqrt{\frac{\mu^3\sigma}{4\pi f^3}}\right\}^2} - 1\right\}m_e \quad (16)$$

Note that the *momentum* $q$ in Eq.(1) can be also produced by an Electric and/or Magnetic field *if the particle has an electric charge Q*.

In that case, combination of *Lorentz's Equation* $\vec{F} = Q\vec{E}_0 + Q\vec{V}\times\vec{B}$ and $\vec{F} = m_g \vec{a}$ (see reference 1, p.78-Eq.(2.05)) gives

$$q' = m_g V = m_g \frac{Q(\vec{E}_0 + \vec{V}\times\vec{B})}{m_g}\Delta t \quad (17)$$

In the particular case of an oscillating EM field ( frequency $f_{osc}$, $\Delta t = 1/f_{osc}$ ) we have

$$q' = \frac{Q(\vec{E}_{osc} + \vec{V}\times\vec{B}_{osc})}{f_{osc}} \quad (18)$$

Thus, the general expression of $q$ in Eq.(1) will be

$$q = \frac{U}{v} + q' = \frac{U}{c}n_r + \frac{Q(\vec{E}_{osc}+\vec{V}\times\vec{B}_{osc})}{f_{osc}} \quad (19)$$

Consequently, if the *Hg* cylinder *rotates* at an *oscillating* magnetic field $B_{osc}$ of frequency $f_{osc}$, perpendicular to the cylinder, then the *total* value of $q$ for the electrons of the *Hg*, according to Eq.(19), will be given by

$$q = \frac{U}{c}n_r + \frac{e(\vec{E}_{osc}+\vec{V}\times\vec{B}_{osc})}{f_{osc}} \quad (20)$$

where $\frac{U}{c}\eta_r$, according to Eq.(16), is given by

$$\frac{U}{c}\eta_r = \frac{e^4 B^2 V^2}{6\pi m_e^2 c(1-V^2/c^2)}\sqrt{\frac{\mu^3\sigma}{4\pi f^3}}$$

where

$$\vec{V} = \vec{V}_{\omega_c} + \vec{V}_{osc}$$

In the equation above $V_{\omega_c} = \omega_c R$ and $V_{osc}$ can be calculated by means of the well-known equations of the *Ohm's vectorial Law* : $\vec{J} = \sigma\vec{E}$ and $\vec{J} = \rho_m \vec{V}$ ($J$ is the current density, in A/m$^2$ ; $\rho_m$ and $V$ are respectively, the density (C/m$^3$) and the velocity of charge carriers). Thus we can write

$$V_{osc} = \left(\frac{\sigma}{\rho_m}\right)E_{osc} = \left(\frac{\sigma}{\rho_m}\right)vB_{osc} =$$
$$= \left(\frac{\sigma}{\rho_m}\right)B_{osc}\sqrt{\frac{2\omega_{osc}}{\mu\sigma}} = B_{osc}\sqrt{\frac{4\pi f_{osc}\sigma}{\mu\rho_m^2}} \quad (21)$$

Thus the Eq.(20) can be rewritten in the following form:

$$q = \frac{e^4 B_{osc}^2 V_{osc}^2}{6\pi m_e^2 c^2 \left(1 - V_{osc}^2/c^2\right)} \sqrt{\frac{\mu^3 \sigma}{4\pi f^3}} + \frac{e\left(E_{osc} + \omega_c R B_{osc} + V_{osc} B_{osc}\right)}{f_{osc}} \quad (22)$$

Where $E_{osc} = v B_{osc}$ ( $v$ given by the Eq.(3), $v = \sqrt{\frac{2\omega_{osc}}{\mu\sigma}}$ ).

By the substitution of Eq.(22) into Eq.(1), we obtain

$$m_{ge} = m_e - 2\left\{\sqrt{1 + \left[X + \frac{e(E_{osc} + \omega_c R B_{osc} + V_{osc} B_{osc})}{(m_e c f_{osc})}\right]^2} - 1\right\} m_e \quad (23)$$

where

$$X = \frac{e^4 B_{osc}^2 V_{osc}^2}{6\pi m_e^3 c^2 \left(1 - V_{osc}^2/c^2\right)} \sqrt{\frac{\mu^3 \sigma}{4\pi f^3}}$$

For $H_g$ at the superconducting state we can take $\mu \cong \mu_0$ ; $\sigma \cong 10^{23} S/m$ and $\rho_m \cong 10^{13} C/m^3$. Thus, when the $Hg$ cylinder is rotating at angular frequency $\omega_c \cong 5000 rpm$, within a magnetic field $B_{osc} \cong 0.1T$ of frequency $f_{osc} \cong 10 Mhz$, a point at distance $R = 10 cm$ ( *average* radius of the cylinder ) from the rotating axis has tangential velocity $V_{\omega_c} = \omega_c R \cong 52 m/s$, and consequently the gravitational masses of the electrons at this distance are then

$$m_{ge} \cong -160.70 m_e$$

We can assume this value as the *average* gravitational mass of the electrons. Thus, the *total average* gravitational mass can be written as follows

$$m_{g(Hg)} = \left[\frac{-160.70 m_e + m_p + m_n}{m_e + m_p + m_n}\right] m_{i(Hg)} \cong$$
$$\cong 0.96 m_{i(Hg)}$$

This means that the *total gravitational mass* of the $Hg$ decreases of approximately 4% percent.

In our opinion, this way *Podkletnov's effect*[7] may be understood.

When the $Hg$ cylinder isn't rotating ($\omega_c = 0$) the Eq.(23) reduces to

$$m_{ge} = m_e - 2\left\{\sqrt{1 + \left[\frac{e}{m_e c}\left(\frac{E_{osc}}{f_{osc}} + \frac{V_{osc} B_{osc}}{f_{osc}}\right)\right]^2} - 1\right\} m_e$$
$$= m_e - 2\left\{\sqrt{1 + \left[\frac{e B_{osc}}{m_e c}\left(\sqrt{\frac{4\pi}{\mu\sigma f_{osc}}} + \frac{V_{osc}}{f_{osc}}\right)\right]^2} - 1\right\} m_e \quad (24)$$

By the substitution of Eq.(21) into Eq.(24) we obtain

$$m_{ge} = m_e - 2\left\{\sqrt{1 + \left[\frac{e B_{osc}}{m_e c}\left(\sqrt{\frac{4\pi}{\mu\sigma f_{osc}}} + B_{osc}\sqrt{\frac{4\pi\sigma}{\mu\rho_m^2 f_{osc}}}\right)\right]^2} - 1\right\} m_e \quad (25)$$

Then, if $B_{osc} \cong 10T$ ; $f_{osc} \cong 10 MHz$, Eq.(25) gives

$$m_{ge} \cong -3700 m_e$$

Thus, the *total average* gravitational mass of the $H_g$ is

$$m_{g(Hg)} = \left[\frac{-3700 m_e + m_p + m_n}{m_e + m_p + m_n}\right] m_{i(Hg)} \cong -0.01 m_{i(Hg)}$$

Again we suggest to check the weights of samples (of different masses and chemical compositions) above the $Hg$. Possibly the samples will *float* above the *Hg* ( the gravitational masses of the samples will be *slightly negative*, due to the *negative* gravitational mass of the *Hg* ).

Let us now consider a *static* ( $\omega_c = 0$ ) *parallel-plate capacitor*, where $d$ is the distance between the plates; $\Delta V_{AC}$ is the applied voltage; $E_{osc} = \Delta V_{AC}/d$ is the external electric field. Inside the dielectric the electric field is $E = \sigma_m/\varepsilon = E_{osc}/\varepsilon_r$ where $\sigma_m$ (in C/m$^2$) is the density of electric charge and $\varepsilon = \varepsilon_r \varepsilon_0$.

Thus the charge $Q$ on each surface of the dielectric is given by $Q = \sigma_m S$ ( $S$ is the area of the surface). Then we have

$$Q = \sigma_m S = (E_{osc}\varepsilon_0)S \quad (26)$$

Within the oscillating field $E_{osc}$ the charge $Q$ (or "charge layer") acquire a *momentum* $q$, according to Eq.(19), given by

$$q = \frac{U}{c}n_r + \frac{Q(E_{osc} + \vec{V}_{osc} \times \vec{B}_{osc})}{f_{osc}} =$$
$$= \frac{U}{c}n_r + \frac{Q(2E_{osc})}{f_{osc}} \quad (27)$$

If $U = 0$ then Eq.(27) reduces to

$$q = \frac{2QE_{osc}}{f_{osc}} = \frac{2E_{osc}^2 \varepsilon_0 S}{f_{osc}} = \frac{2(\Delta V_{AC}/d)^2 \varepsilon_0 S}{f_{osc}} \quad (28)$$

Assuming that in the dielectric of the capacitor there are $N^*$ *layers of dipoles* with thickness $\xi$ approximately equal to the diameter of the atoms, i.e., $N^* = d/\xi \cong 10^{10}d$ then, according to Eq.(1), for $q \gg m_i c$, the gravitational mass $m_g^*$ of each *dipole layer* is

$$m_g^* \cong -2\left(\frac{q}{m_i c}\right)m_i \cong -\frac{2q}{c} \cong$$
$$\cong -4\left(\frac{\Delta V_{AC}}{d}\right)^2 \frac{\varepsilon_0 S}{f_{osc}c} \quad (29)$$

Thus, the total gravitational mass $m_g$ of the dielectric may be written in the following form

$$m_g = N^* m_g^* \cong -4\times 10^{10}\left(\frac{\varepsilon_0 S}{f_{osc}cd}\right)\Delta V_{AC}^2 \quad (30)$$

For example, if we have $\Delta V_{AC} = 50KV; S = 0.01m^2; f_{osc} \cong 10^2 Hz$ and $d = 1mm$
Eq.(30) gives $m_g \cong -0.3kg$

The result above can also be reach by means of the calculation of the gravitational masses of the electrons of the dielectric of the capacitor. Note that the acceleration upon the *electrons* (due to the field $E_{osc}$) is obviously equal to acceleration upon the *electric dipoles* of the dielectric. Consequently, the momentum $q$ for the *electrons* $(q_e)$ and for the *electric dipoles* $(q_{dip})$ are respectively, $q_e = m_e V$ and $q_{dip} = m_{dip} V$. Thus, $q_e = q_{dip}\left(\frac{m_e}{m_{dip}}\right)$.

From Eq.(27), for $U = 0$, we can write

$$q_{dip} = \frac{2Q_{dip}E_{osc}}{f_{osc}} \quad (31)$$

where $Q_{dip}$ is the *dipole electric charge*. Consequently,

$$q_e = \frac{2Q_{dip}E_{osc}}{f_{osc}}\left(\frac{m_e}{m_{dip}}\right) \quad (32)$$

and

$$\frac{q_e}{m_e c} = \frac{2Q_{dip}E_{osc}}{m_e c f_{osc}}\left(\frac{m_e}{m_{dip}}\right) = \frac{2Q_{dip}E_{osc}}{m_{dip}c f_{osc}} \quad (33)$$

By the substitution of Eq.(33) into Eq.(1), we obtain

$$m_{ge} = m_e - 2\left\{\sqrt{1+\left\{\frac{2Q_{dip}E_{osc}}{m_{dip}c f_{osc}}\right\}^2} - 1\right\}m_e \quad (34)$$

Assuming $Q_{dip} \cong 2\times10^{-19}C$; $m_{dip} \cong 1\times10^{-26}kg$ and $\Delta V_{AC} = 50KV$; $d = 1mm$; $f_{osc} \cong 10^2 Hz$ then Eq.(34) gives

$$m_{ge} \cong -133000 m_e$$

Then the total gravitational mass of the dielectric is

$$m_g = \left[\frac{-133000 m_e + m_p + m_n}{m_e + m_p + m_n}\right]m_{i(Hg)} \cong -35 m_i$$

But $m_i = \rho V = \rho S d \cong 10^{-2}kg$ then

$$m_g \cong -0.3 kg$$

Possibly this is the explanation for the *Biefeld-Brown Effect*.